\documentclass{article}

\usepackage{arxiv}

\usepackage[utf8]{inputenc} 
\usepackage[T1]{fontenc}    
\usepackage{hyperref}       
\usepackage{url}            
\usepackage{booktabs}       
\usepackage{amsfonts}       
\usepackage{nicefrac}       
\usepackage{microtype}      
\usepackage{lipsum}		
\usepackage{graphicx}

\title{Constructing a Data Visualization Recommender System}

\author{
  Petra Kubern\'{a}tov\'{a}\ \\
  Leiden Institute of Advanced Computer Science\\
  Leiden University\\
  Leiden, Netherlands\\
  \texttt{pkubernatova@gmail.com} \\
   \And
  Magda Friedjungov\'{a} \\
  Faculty of Information Technology\\
  Czech Technical University in Prague\\
  Prague, Czech Republic\\
  \texttt{magda.friedjungova@fit.cvut.cz} \\
   \And
  Max van Duijn\\
  Leiden Institute of Advanced Computer Science\\
  Leiden University\\
  Leiden, Netherlands\\
  \texttt{m.j.van.duijn@liacs.leidenuniv.nl} \\
}

\begin{document}
\maketitle

\begin{abstract}
Choosing a suitable visualization for data is a difficult task. Current data visualization recommender systems exist to aid in choosing a visualization, yet suffer from issues such as low accessibility and indecisiveness. In this study, we first define a step-by-step guide on how to build a data visualization recommender system. We then use this guide to create a model for a data visualization recommender system for non-experts that aims to resolve the issues of current solutions. The result is a question-based model that uses a decision tree and a data visualization classification hierarchy in order to recommend a visualization. Furthermore, it incorporates both task-driven and data characteristics-driven perspectives, whereas existing solutions seem to either convolute these or focus on one of the two exclusively. Based on testing against existing solutions, it is shown that the new model reaches similar results while being simpler, clearer, more versatile, extendable and transparent. The presented guide can be used as a manual for anyone building a data visualization recommender system. The resulting model can be applied in the development of new data visualization software or as part of a learning tool.
\end{abstract}


\section{Introduction}
In recent years, we have been witnesses to the rise of big data as one of the key topics of computer science. With huge amounts of data being generated every second of every day, it has become a necessity to focus on its storage, analysis and presentation. Data visualization is aiding us in the presentation phase. By definition, it is the representation of information in a visual form, such as a chart, diagram or picture. It can find its place in a variety of areas such as art, marketing, social relations and scientific research. There were over 300 visualization types available at the time of writing this paper \cite{d3}. This growing number makes the choice of choosing a suitable data visualization very difficult. Especially if one is not particularly skilled in the area. Thankfully, data visualization recommender systems exist to help us with this complicated task. 

This paper is an extended version of \cite{nevim} which was presented at the DATA 2018 conference. The main contribution of this extended version lies in including a step-by-step guide for building a data visualization recommender system. We also present a more extensive existing solution study. We further clarify the process of constructing our model and also elaborate on ways of evaluating and implementing it. Section \ref{context} places data visualization recommender systems in the context of data science. Section \ref{process} introduces our step-by-step guide to building a data visualization recommender system. In Sections 4-10 we go through the individual steps and build our very own data visualization recommender system while taking measures to make it well suited for non-expert users. We define a 'non-expert user' as someone without professional or specialized knowledge of data visualization. We thus include both complete beginners and users who have general knowledge of data visualization types (e.g. bar charts, pie charts, scatter plots) but have no professional experience in the fields of data science and data communication. We want to see if we can make adjustments that make a system more suitable for non-expert users while maintaining effectiveness (still clearly distinguishing the data visualizations from each other) and performance (recommending the most suitable visualization type). We draw conclusions in Section \ref{conclusion} and set an agenda for future work in Section \ref{future}.

\section{Context} \label{context}
\subsection{Data science}

\begin{figure}
\centering
\includegraphics[width=0.6\textwidth]{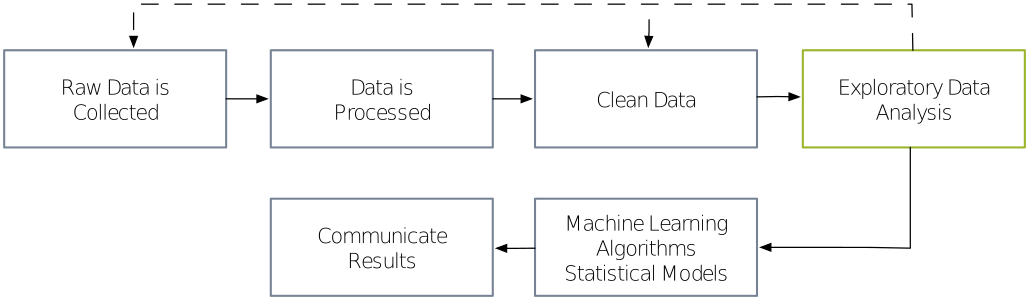}
\caption{ The data science process \cite{oneil} } \label{datascience}
\end{figure}

\noindent Data science plays an important role in scientific research, as it aids us in collecting, organizing, and interpreting data, so that it can be transformed into valuable knowledge. Figure \ref{datascience} shows a simplified diagram of the data science process as described by O\textsc{\char13}Neil and Schutt \cite{oneil}. This diagram is helpful in demarcating the research objectives of this paper. According to O\textsc{\char13}Neil and Schutt, first, real world raw data is collected, processed and cleaned through a process called data munging. Then exploratory data analysis (EDA) follows, during which we might find that we need to collect more data or dedicate more time to cleaning and organizing the current dataset. When finished with EDA, we may use machine learning algorithms, statistical models and data visualization techniques, depending on the type of problem we are trying to solve. Finally, results can be communicated \cite{oneil}.

Our focus here is on the part of the process concerning exploratory data analysis or EDA. EDA uses a variety of statistical techniques, principles of machine learning, but also, crucially, the data visualization techniques we study in this paper. Please note that data visualization can also be a part of the “Communicate Results” stage of the data science process. There is a thin line between data visualizations made for exploration and ones made for explanation, as most exploratory data visualizations also contain some level of explanation and vice-versa.

\subsection{Exploratory data analysis}

Exploratory data analysis (EDA) is not only a critical part of the data science process, it is also a kind of philosophy. You are aiming to understand the data and its shape and connect your understanding of the process that collected the data with the data itself. EDA helps with suggesting hypotheses to test, evaluating the quality of the data, identifying potential need for further collection or cleaning, supporting the selection of appropriate models and techniques and, most importantly for the context of this study, it helps find interesting insights in your data \cite{tukey}. 

\subsection{Data visualization}

There are many definitions of the term data visualization. The one used in this study is: data visualization is the representation and presentation of data to facilitate understanding. According to Kirk, our eye and mind are not equipped to easily translate the textual and numeric values of raw data into quantitative and qualitative meaning. "We can look at the data, but we cannot understand it. To truly understand the data, we need to see it in a different kind of form. A visual form." \cite{kirk}

Illinsky and Steele describe data visualization as a very powerful tool for identifying patterns, communicating relationships and meaning, inspiring new questions, identifying sub-problems, identifying trends and outliers, discovering or searching for interesting or specific data points \cite{illinsky}.

Tamara Munzner made a 3-step model for data visualization design. According to this model, we first need to decide what we want to show. Secondly, we need to motivate why we want to show it. Finally, we need to decide how we are going to show it \cite{munzner}. There are many different types of data visualizations to help us with the third step. However, the challenge remains in choosing the most suitable one. Data visualization recommender systems were made to help with this difficult task. We find that the WHAT and the WHY greatly influence the HOW, thus we aim to build a system that reflects all three aspects of the data visualization design process in some way.

\subsection{Data visualization recommender systems}

Within this study we define data visualization recommender systems as tools that seek to recommend visualizations which highlight features of interest in data. This definition is based on combining common aspects of definitions in existing work.

While the output of data visualization recommender systems is always a recommendation for data visualization types in some shape or form, the input can differ. It can be, for example, just the data itself, a specification of goals or the specification of aesthetic preferences. The type of input affects the type of recommendation strategy used and consequently the type of the recommender system. Kaur and Owonibi distinguish 4 types of recommender systems \cite{kaur}:
\begin{itemize}
\item \textbf{Data Characteristics Oriented:} These systems recommend visualizations based on data characteristics. 
\item \textbf{Task Oriented:} These systems recommend visualizations based on representational goals as well as data characteristics.
\item \textbf{Domain Knowledge Oriented:} These systems improve the visualization recommendation process with domain knowledge.
\item \textbf{User Preferences Oriented:} These systems gather information about the user presentation goals and preferences through user interaction with the visualization system. 
\end{itemize}

The line between different categories of recommendation systems is rather thin and some systems can have ambiguous classifications, as will be discussed.

\section{Building a data visualization recommender system} \label{process}

\subsection{Previous work overview}

In any type of research, it is important to explore the state of the art of the corresponding field first. We will examine the area of data visualization recommender systems and its history. We are going to look at current solutions and identify which aspects could be used in our own system. Secondarily, the previous work overview will also help us determine which solutions are most suitable to be used while testing our system later on.

\subsection{Exploratory survey}

Next, it is crucial to get the potential users involved. We will create and run an exploratory survey among different data science communities on Facebook and LinkedIn. We focus on data science communities, because it will ensure that our respondents will have some sort of familiarity with data science and its terminology. We made this choice for convenience and time management reasons. The survey should aid us in making decisions about our system. As we want to explore options of making the system easy to use for non-experts, it is important to have a clear definition of a non-expert. The survey will help us in this specification as well.

\subsection{Forming requirements}

The results of both the previous work overview and the survey will help us formulate requirements which our system should fulfill. These will serve as good evaluation criteria for our system later on.

\subsection{Constructing a model}

Once the requirements are formulated, we will begin constructing a model for our recommender system. First we will have to decide which form the model will take - what the base structure will be. We will then establish the different components of the structure. Finally, we will combine it all together.

\subsection{Testing the model}

At this point we will take time to test the model that we have constructed. We will perform two tests. The first one will focus on determining whether the model is able to produce results similar to existing solutions. The second test focuses on testing the extendibility of the model by adding a new type of visualization.

\subsection{Evaluating the model}

We will evaluate the model according to the results of the tests and also comment on how it manages to fulfill the requirements we have set. We will most probably have to make adjustments to the model.

\subsection{Implementing the model}

Finally, we will implement the model we have created and make it come to life as a data visualization recommender system.

\section{Previous work overview} \label{building}

While performing the existing solutions study, we have discovered that the amount of works in the area focusing on domain knowledge oriented and user preferences oriented data visualization recommender systems is not sufficient. We made the decision to focus on data characteristics oriented and task oriented systems. As we stated earlier, there is a thin line between the different types of recommendation systems, so in some solutions you might still see elements of the two types of recommender systems we do not explicitly focus on.

\subsection{Data Characteristics Oriented systems}

Systems based on data characteristics aim to improve the understanding of the data, of different relationships that exist within the data and of procedures to represent them. Some of the following tools and techniques are not recommendation systems per se but they were a crucial part of the history of this field and foundations for other recommender systems stated, thus we feel it is appropriate to list them as well.

\subsubsection{BHARAT}

was the first system that proposed rules for determining which type of visualization is appropriate for certain data attributes \cite{bharat}. As this work was written in 1981, the set of possible visualizations was not as varied as it is today. The system incorporated only the line, pie and bar charts and was based on a very simple design algorithm. If the function was continuous, a line chart was recommended. If the user indicated that the range sets could be summed up to a meaningful total, a pie chart was recommended and bar charts were recommended in all the remaining cases. Even though this system would now be considered very basic, it served as the foundation for other systems that followed.

\subsubsection{APT}

In 1986, Mackinlay proposed to formalize and codify the graphical design specification to automate the graphics generation process \cite{apt}. His work is based on the work of Joseph Bertin, who came up with a semiology of graphics, where he specified visual variables such as position, size, value, color, orientation etc. and classified them according to which features they communicate best. For example, the shape variable is best used to show differences and similarities between objects. Mackinlay codified Bertin’s semiology into algebraic operators that were used to search for effective presentations of information. He based his findings on the principals of expressiveness and effectiveness. Expressiveness is the idea that graphical presentations are actually sentences of graphical languages that have precise syntactic and semantic definitions, while effectiveness refers to how accurately these presentations are perceived. He aimed to develop a list of graphical languages that can be filtered with the expressiveness criteria and ordered with the effectiveness criteria for each input. He would take the encoding technique and formalize it with primitive graphical language (which data visualizations can show this), then he would order these primitive graphical languages using the effectiveness principle (how accurately perceived they are). APT\textsc{\char13}s architecture was focused on how to communicate graphically rather than on what to say. Casner extended this work by comparing design alternatives via a measure of the work that was required to read presentations, depending on the task \cite{casner}. Roth and Mattis added additional types of visualizations \cite{roth}.

\subsubsection{VizQL(Visual Query Language)}
In 2003, Hanrahan revised Mackinglay\textsc{\char13}s specifications into a declarative visual language known as VizQL \cite{vizql}. It is a formal language for describing tables, charts, graphs, maps, time series and tables of visualizations. The language is capable of translating actions into a database query and then expressing the response graphically.

\subsubsection{Tableau and its Show Me Feature}

The introduction of Tableau was a real milestone in the world of data visualization tools. Due to the simple user interface, even inexperienced users could create data visualizations. It was created when Stolte, together with Hanrahan and Chabot, decided to commercialize a system called Polaris \cite{stolte} under the name Tableau Software. In 2007 Tableau introduced a feature called Show Me \cite{mackinlay2}. The Show Me functionality takes advantage of VizQL to automatically present data. At the heart of this feature is a data characteristics-oriented recommendation system. The user selects the data attributes that interest him and Tableau recommends a suitable visualization. Tableau determines the proper visualization type to use by looking at the types of attributes in the data. Each visualization requires specific attribute types to be present before it can be recommended. For example, a scatter plot requires 2 to 4 quantitative attributes. Furthermore, the system also ranks every visualization on familiarity and design best practices. Finally, it recommends the highest-ranked eligible visualization. Mackinlay and his team have also performed interesting user tests with the Show Me feature. They found that the Show Me feature is being used (very) modestly by skilled users (i.e. in only 5.6\% of cases).

Tableau inspired us by it’s simple user interface which is suitable for non-experts, reminding us that our model should enable a simple user interface implementation. Furthermore, we make use of their classification of data visualizations based on design best practices and familiarity as well as the conditions that the data must fulfill for a specific data visualization to be chosen. The fact that Tableau is so widely used and that a demo version is freely accessible determined it suitable for use in our tests.

\subsubsection{ManyEyes}
Viegas et al. created the first known public website where users may upload data and create interactive visualizations collaboratively: ManyEyes \cite{manyeyes}. The tool was created for non-experts, as Viegas et al. wanted to make a tool that was accessible for anyone regardless of prior knowledge and training. Design choices were made to reflect the effort to find a balance between powerful data-analysis capabilities and accessibility to the non-expert visualization user. The visualizations were created by matching a dataset with one of the 13 types of data visualizations implemented in the tool. To set up this matching, the visualization components needed to be able to express its data needs in a precise manner. They divided the data visualizations into groups by data schemas. A data schema could be, for example, “single column textual data”. Thus, a bar chart was described as “single column textual data and more than one numerical value”. The dataset and produced visualization could then be shared with others for comments, feedback and improvement \cite{manyeyes}. However, the tool closed down in 2015.  

ManyEyes taught us that the way to attract non-expert users is to make the application resulting from our model as accessible as possible. This means that our model could be suited to web-based implementations.

\subsubsection{Watson Analytics}
Since 2014, IBM have been developing a tool called Watson Analytics \cite{watson}. It carries the same name as another successful IBM project - the Watson supercomputer, which combines artificial intelligence and sophisticated analytical software to perform as a “question-answering” system. In 2011, it famously defeated top-ranked players in a game of Jeopardy!. Similarly to the Watson supercomputer, Watson Analytics uses principles of machine learning and natural language processing to recommend users either questions they can ask about their data, or a specific visualization. However, IBM has not revealed what values or attributes are used by the recommendation system to select a visualization.

Watson Analytics reminded us that the structure of our model should be variable enough to be suitable for implementing machine learning and artificial intelligence techniques on it for the model to possibly improve itself. A demo version of the system is freely available, so we use it in our tests. 

\subsubsection{VizDeck}

In 2012 Key et al. developed a tool called VizDeck \cite{vizdeck}. The web-based tool recommends visualizations based on statistical properties of the data. It adopts a card game metaphor to organize multiple visualizations into an interactive visual dashboard application.  Vizdeck was created as Key et al. found that scientists were not able to self-train quickly in more sophisticated tools such as Tableau. The tool supports scatter plots, histograms, bar charts, pie charts, timeseries plots, line plots and maps. 

Based on the statistical properties of the underlying dataset, VizDeck generates a “hand” of ranked visualizations and the user chooses which “cards” to keep and put into a dashboard and which to discard. Through this, the system learns which visualizations are appropriate for a given dataset and improves the quality of the “hand” dealt to future users. For the actual recommendation system part of the tool, they trained a model of visualization quality that relates statistical features of the dataset to particular visualizations. As far as we know VizDeck was never actually deployed and remained at the testing phase. 

VizDeck again inspired us to think about the possibility of our model being self-improving and educative.

\subsubsection{Microsoft Excel’s Recommended Charts Feature} In the 2013 release of Microsoft Excel, a new feature called Recommended Charts was introduced. The user can select the data they want to visualize and Excel recommends a suitable visualization \cite{excel}. However, Microsoft does not share exactly how this process is carried out, making it less suitable as a source of inspiration. We use Microsoft Excel to test our model, because it is accessible. 

\subsubsection{SEEDB}

In 2015 Vartak et al. proposed an engine called SEEDB \cite{seedb}. They judge the interestingness of a visualization based on the following theory: a visualization is likely to be interesting if it displays large deviations from some reference (e.g. another dataset, historical data, or the rest of the data). This helps them identify the most interesting visualizations from a large set of potential visualizations. They identified that there are more aspects that determine the interestingness of a visualization, such as aesthetics, user preference, metadata and user tasks. A full-fledged visualization recommendation system should take into account a combination of these aspects. A major disadvantage of SEEDB is that it only uses variations of bar charts and line charts. As far as we know SEEDB was never deployed.

SEEDB made us think about having multiple views in our model from different interestingness perspectives, because we want our model to be full-fledged, as they describe.

\subsubsection{Voyager}

In 2016, Wongsuphasawat et al. developed a visualization recommendation web application called Voyager \cite{voyager}, based on the Compass recommendation engine \cite{compass} and a high-level specification language called Vega-lite \cite{vegalite}. It couples browsing with visualization recommendation to support exploration of multivariate, tabular data. First, Compass selects variables by taking user-selected variable sets and suggesting additional variables. It then applies data transformations (e.g. aggregation or binning) to produce a set of derived data tables. For each data table, it designs encodings based on expressiveness and effectiveness criteria and prunes visually similar results. The user then includes or excludes different variables to focus on a particular set of variables that are interesting. Voyager is a tool which is freely available online, which makes it suitable for use in our tests.

\subsubsection{Google Sheets}
Google Sheets \cite{googlesheets} is a tool which allows users to create, edit and share spreadsheets. It was introduced in 2007 and is very similar to Microsoft Excel. In June of 2017, the tool was extended with the Explore Feature, which helps with automatic chart building and data visualization. It uses elements of artificial intelligence and natural language processing to recommend users questions they might want to ask about their data, as well as recommending data visualizations that best suit their data. In the documentation for this feature, Google specifies each of the included data visualizations using functions and conditions that have to be fulfilled in order for that particular data visualization to be recommended. However, it does not reveal exactly how it chooses the most suitable data visualization, because a couple of visualizations have the same conditions. We make use of the classification of data visualizations presented in Google Sheets and thanks to its accessibility online, we use it in our tests.

\subsection{Task Oriented systems}

Task-oriented systems aim to design different techniques to infer the representational goal or a user’s intentions. In 1990, Roth and Mattis were the first to identify different domain-independent information seeking goals, such as comparison, distribution, correlation etc. \cite{roth}.  Also in 1990, Wehrend and Lewis proposed a classification scheme based on sets of representational goals \cite{wehrend}. It was in the form of a 2D matrix where the columns were data attributes, the rows representational goals and the cells data visualizations. To find a visualization, the user had to divide the problem into subproblems, until for each subproblem it was possible to find an entry in the matrix. A representation for the original complex problem could then be found by combining the candidate representation methods for the subproblems. Unfortunately, the complete matrix was not published so it is unknown which specific types of data visualizations were included.

\subsubsection{BOZ}

BOZ is an automated graphic design and presentation tool that designs graphics based on an analysis of the task which a graphic is intended to support \cite{casner}. The system analyzes a logical description of a task to be performed and designs an equivalent perceptual task. BOZ produces a graphic along with a perceptual procedure describing how to use the graphic to complete the task. It is able to design different presentations of the same information customized to the requirements of different tasks. 

The BOZ system reminded us that the difference between a suitable and non-suitable data visualization could also lie in the way that humans perceive them. For example, a pie chart is generally considered not suitable, as humans have difficulty judging the size of angles accurately. A bar chart is more suitable for the task. 

\subsubsection{IMPROVISE}

In the previous studies, the user task list was manually created. However, in 1998, Zhou and Feiner introduced advanced linguistic techniques to automate the derivation of the user task from a natural language query \cite{improvise}. They introduced a visual task taxonomy to automate the process of gaining presentation intents from the text. The taxonomy interfaces between high level tasks that can be accomplished by low level visualization techniques. For example, the visual task “Focus” implies that visual techniques such as “Enlarge” or “Highlight” could be used. This taxonomy is implemented in IMPROVISE. 

An example of an IMPROVISE use case is presenting an overview of a hospital patient's information to a nurse. To achieve this goal, it constructs a structure diagram that organizes various information (e.g. IV lines) around a core component (the patient\textsc{\char13}s body). In a top-down design manner, IMPROVISE first creates an ‘empty’ structure diagram and then populates it with components by partitioning and encoding the patient information into different groups.

\subsubsection{HARVEST}

In 2009, Gotz and Wen introduced a novel behavior-driven approach \cite{gotz}. Instead of needing explicit task descriptions, they use implicit task information obtained by monitoring users' behavior to make recommendation more effective. The Behavior-Driven Visualization Recommendation (BVDR) approach has two phases. In the first phase of BDVR, they detect four predefined patterns from user activity. In the second phase, they feed the detected patterns into a recommendation algorithm, which infers user intent in terms of common visual tasks (e.g. comparison) and suggests visualizations that better support the user's needs. The inferred visual task is used together with the properties of the data to retrieve a list of potentially useful visual metaphors from a visualization example corpus made by Zhou and Chen \cite{zhou2}. It contains over 300 examples from a wide variety of sources. Unfortunately, we were not able to access this corpus.

The conclusions made from HARVEST gave us the idea to provide explanations why a certain data visualization was recommended to enhance the educative aspect of our model. 

\subsubsection{DataSlicer}

A recent study by Alborzi et al. takes yet another approach \cite{dataslicer}. The authors' hypothesis is that for many data sets and common analysis tasks, there are relatively few “data slices” that result in effective visualizations. Data slices are different subsets of data. Their objective is to improve the user experience by suggesting data slices that, when visualized, present correct solutions to the user’s task in an effective way. At any given time in working on the task, users may ask the system to suggest visualizations that would be useful for solving the task. A data slice is considered interesting if past users spent a considerable amount of time looking at its visualization. They first developed a framework which captures exemplary data slices for a user task, explores and parses visual-exploration sequences into a format that makes them distinct and easy to compare. Then they developed a recommendation system, DataSlicer, that matches a "currently viewed" data slice with the most promising "next effective" data slices for the given exploration task. In user tests, DataSlicer significantly improved both the accuracy and speed for identifying spatial outliers, data outliers, outlier patterns and general trends. The system quickly predicted what a participant was searching for based on their initial operations, then presented recommendations that allowed the participants to transform the data, leading them to desired solutions. 

The system is interesting, because it deals with the problem of efficiently leading casual or inexperienced users to visualizations of the data that summarize in an effective and prominent way the data points of interest for the user’s exploratory-analysis task. The authors do not specify exactly which tasks they include in their system.  

All in all, we identify some pitfalls of the existing systems. Such as them not being accessible enough, too complicated, too formal and too secretive when it comes to their recommendation process. The biggest pitfall is that the result of their recommendation process is most commonly a set of data visualizations, which, in our opinion, leaves the users a bit further than they started, but still nowhere, because they still have to choose the most suitable visualization. The possibilities have been narrowed, but a decision still must be made. We hope to avoid these pitfalls within our model.

We establish that we are going to test our model against the solutions available to us. This means Tableau, Watson Analytics, Excel, Voyager and Google Sheets. Please note that we are going to compare against the recommendation system features of the tools, not the tools as a whole.

\section{Exploratory Survey}

\subsection{Participants}

In total, we gathered 88 valid responses (\textit{n=88}). Out of the 88 respondents, 78\% (\textit{n=69}) were male and 22\% (\textit{n=19}) female. The average age was 29.86 years. We asked the respondents to indicate their knowledge level on a scale of 1 to 10, 1 being beginner and 10 being expert. The average knowledge level was 5.70. We opted to divide the scale into three ranges in the following way: 1-3 are beginners, 4-7 are non-experts and 8-10 are experts. According to our ranges we had 26\% (\textit{n=23}) beginner level, 44\% non-expert (\textit{n=39}) level and 30\% (\textit{n=26}) expert level respondents. 

\subsection{Questions}

Questions in our survey included:
\begin{itemize}
\item How long have you been working in a data visualization related field? 
\item Which software do you mostly use to create your data visualizations?
\item What are your top 3 most used visualization techniques? (e.g. bar chart, scatterplot, line chart, treemap...)
\item What is your main goal when you make data visualizations?
\item Which of the following tasks do you usually perform using your data visualization?
\item Do you know any data visualization recommender systems?
\end{itemize}

\subsection{Results}

The results of our survey can be summarised as follows:

\begin{itemize}
\item For all groups, the main purpose of making data visualizations was for analysis (65\% of beginners, 64\% of non-experts, 58\% of experts).
\item All types of users choose data visualizations mainly according to: the characteristics of their data (57\% of beginners, 62\% of non-experts, 65\% of experts) and the tasks that they want to perform (48\% of beginners, 51\% of non-experts, 62\% of experts).
\item For all groups, the two most used visualizations are bar charts (17\% of beginners, 38\% of non-experts, 35\% of experts) and scatter plots (43\% of beginners, 26\% of non-experts, 31\% of experts).
\item All groups were mostly unable to name an existing data visualization recommendation system (0\% able vs. 100\% unable for beginners, 5\% able vs. 95\% unable for non-experts and 4\% able vs. 96\% unable for experts).
\item All groups would be willing to use a data visualization recommendation system, although experts were less willing than beginners and non-experts (100\% willing vs. 0\% not willing for beginners, 87\% willing vs. 13\% not willing for non-experts and 77\% willing vs. 23\% not willing for experts).
\end{itemize}

\noindent The most crucial finding that we made from the results of our survey was, that the approaches of beginners, non-expert and expert users do not differ enough for us to be able to clearly determine what specific features a non-expert user would need in a data visualization recommender system model. This was an unexpected result for us.

\section{Forming requirements} \label{requirements}

Based on research of previous approaches to our problem and the results of our survey, we have identified the following requirements which our model should fulfill:

\begin{enumerate}
\item \textbf{Simplicity} - The model should be simple, it must have good flow and a very straightforward base structure.
\item \textbf{Clarity} - We aim for the result of our recommendation system to be one data visualization. Not a set, like in some current tools. This means that the underlying classification hierarchy of data visualizations must be clear and unambiguous. 
\item \textbf{Versatility} - We want our model to combine different kinds of recommendation systems. From our survey we learn that when users select a suitable data visualization type, they do so based on the characteristics of their data and the tasks they want to perform. Based on this we incorporate a data characteristics-oriented and task-oriented approach. Furthermore, we want our model to be easily implemented in different programming languages and environments. 
\item \textbf{Extendibility} - Our aim is for our model to be easily extendable. We want the process of adding visualizations into the model to be simple. We want it to be a useful “skeleton” which can be easily extended to include automatic visualizations etc.
\item \textbf{Education} - We want our model to not only function as a recommender system, but also as a learning tool. 
\item \textbf{Transparency} - Once we recommend a visualization, we want the users to see, why that particular visualization was recommended, meaning that the path to a visualization recommendation through our model has to be retraceable. 
\item \textbf{Self-learning} - We want our model to be able to improve itself. This means, amongst other things, that it should be machine learning friendly.
\item \textbf{Competitiveness} - We want our model to still produce results which are comparable to results from other systems.
\end{enumerate}

\section{Constructing a model}

\subsection{Base structure}

Since the aim of our model is to help a user \textit{decide} which data visualization to use, the obvious choice seemed to be to use the structure of decision trees. A decision tree has four main parts: a root node, internal nodes, leaf nodes and branches. Decision trees can help uncover unknown alternative solutions to a problem and they are well suited for machine learning methods.

Once we determined that the decision tree was a possible base structure, we needed to specify what our root node, internal nodes, leaf nodes and branches would be. It was clear right away that the leaf nodes would be the different types of data visualizations since that was the outcome that we wanted to achieve. The root node, internal nodes and branches are inspired by a question-based game that you might have played in your childhood called "21 Questions". In this game, one player thinks of a character and the other players must guess who it is using at most 21 questions. The questions can only be answered 'yes' or 'no'. In our instance, the character the player is thinking of is the leaf node (data visualization), the questions are both internal nodes and the root node and the 'yes' and 'no' answers are branches.

\subsection{Root and internal nodes}

As previously stated, the internal nodes and root node of our model are questions. Just like the "21 Questions" game, each question should eliminate a lot of possible characters (data visualizations). At the same time, the questions have to be understandable (even for non-experts). It became clear that we had to construct questions, which would possess the ability to clearly distinguish different types of data visualizations. The subjects of these questions must then be features that distinguish the different data visualizations from each other. We call these features "distinguishing features". The key to solving this problem is to base the questions on a clear classification hierarchy. As far as we know, there is no one specific classification hierarchy of data visualizations which would be used globally. We had to put together a classification of our own. We decided to research different methods of classification and combine them together. This was a very time-consuming process. We went through a total of 19 books \cite{oneil,kirk,illinsky,munzner,bharat,evergreen,yau1,yau2,heer,hardin,yuk,brath,borner,telea,borner2,ware1,ware2,stacey,hinderman} and for each one, we constructed a diagram representing the classification that was described in the text. We considered the possibility of automating this process, but that would provide enough material to cover a whole other paper.

We examined the classification hierarchies from books together with hierarchies available from web resources and existing tools. We also made note of any advantages or disadvantages of a specific data visualization, if they were listed. For example in several sources \cite{oneil,kirk,illinsky} the authors stated that the pie chart is not suitable for when you have more than 7 parts. The advantages and disadvantages reflected features of the data visualizations that could determine whether they are candidates for recommendation or not, so they are key for the final model. 

We identified that there are two basic views that the classifications incorporate. The first one is a view from the perspective of the task the user wants to perform. The second is a view from the perspective of the characteristics of the data the user has available. This is in line with data characteristics and task oriented recommendation systems \cite{kaur}. 

We have identified a prominent issue in the classification hierarchies: they mix different views into one without making a clear distinction between them. To avoid this issue, we have selected the root node of our model to be a question which would distinguish between two views. The first view is from a task-based perspective and it uses the representational goal or user's intentions behind visualizing the data to recommend a suitable visualization. The second view is from a data-driven perspective, where a visualization recommendation is made based on gathering information about the user's data. The root node of our model is a question asking "Do you know what your main task is?" If the user answers "Yes", he is taken in to the task-based branch. If he answers "No", he is taken straight into the data characteristics-based branch.
 
Once we established the root node, we had to specify the internal nodes. Based on the findings we made in previous paragraphs, we have established a list of distinguishing features and their hierarchy. We present a part of the hierarchy as an example:

\begin{itemize}
\item Suitability for a specific task
\begin{itemize}
\item Comparing
\begin{itemize}
    \item Over time
    \item Quantities
    \item Proportions
    \item Other
\end{itemize}
\item Analyzing
\begin{itemize}
    \item Trends
    \item Correlations
    \item Distribution
    \item Patterns
    \item Clusters
\end{itemize}
\end{itemize}
\end{itemize}

\noindent Based on the distinguishing features, we have constructed questions that ask whether that feature is present or not. For example, for the distinguishing features stated above:

\begin{enumerate}
\item Is your main task to compare over time?
\item Is your main task to compare quantities?
\item Is your main task to compare proportions?
\item Is your main task to compare something else?
\item Is your main task to analyze trends?
\item Is your main task to analyze correlations?
\item Is your main task to analyze distribution?
\item Is your main task to analyze patterns?
\item Is your main task to analyze clusters?
\end{enumerate}

\subsection{Leaf nodes}

The main challenge in this part of the process was to decide which of the more than 300 types of data visualizations available \cite{d3} to include in the initial version of our model. We took a rather quantitative approach to the problem. We went through all the different classification hierarchies we constructed previously and extracted a list of the data visualizations that occur. We removed duplicates (different names for the same visualization, different layouts of the same visualization) and we counted how many times each data visualization occurred. The ones that occurred 5 times or more were included in our final model. The final list contains 29 data visualizations. Since one of our requirements for the final model is easy extendibility, we feel that 29 data visualizations are appropriate for the initial model. The list includes: Bar Chart, Pie Chart, Bubble Chart, Cartogram, Radar Plot, Scatter Plot, Scatter Plot Matrix, Slope Graph, Heat Map, Histogram, Line Chart, Tree Map, Network, Stacked Line Chart etc.

\subsection{Final model}

We classified each of our leaf nodes (data visualizations) using the distinguishing features we constructed previously. For each data visualization, we answered the set of questions relating to the distinguishing features. This revealed how the questions have to be answered in order to get to a certain data visualization.

We then combined all the classifications together to construct the final model\footnote{The whole model can be viewed at a website dedicated to this research project: http://www.datavisguide.com}. The model has 107 internal nodes and 105 leaf nodes. It always results in a recommendation. If no other suitable visualization is found, we recommend to use a table by default. Other data visualization recommender systems such as Tableau adopt this behavior as well.

\section{Testing the model}

\subsection{Competitiveness test}

The aim of this test was to determine whether our model was able to compete with existing systems in terms of similarity of solutions. We obtained 10 different test data sets with various features (See Table \ref{results}). The data sets were preprocessed to remove invalid entries and to ensure that all the attributes were of the correct data type. 

For each data set, we formulated an example question that a potential user is aiming to answer. This was done in order to determine which attributes of the data would be used in the recommendation procedure. Most existing tools require the user to select the specific attributes that they want to use for their data visualization. By specifying these for each data set we attempt to mimic this behavior. Table \ref{results} shows the data sets along with their descriptions.

\begin{table*}[ht]
  \caption{Results of the competitiveness test, our model is labeled NEViM \cite{nevim}}
  \label{results}
  \includegraphics[width=\textwidth]{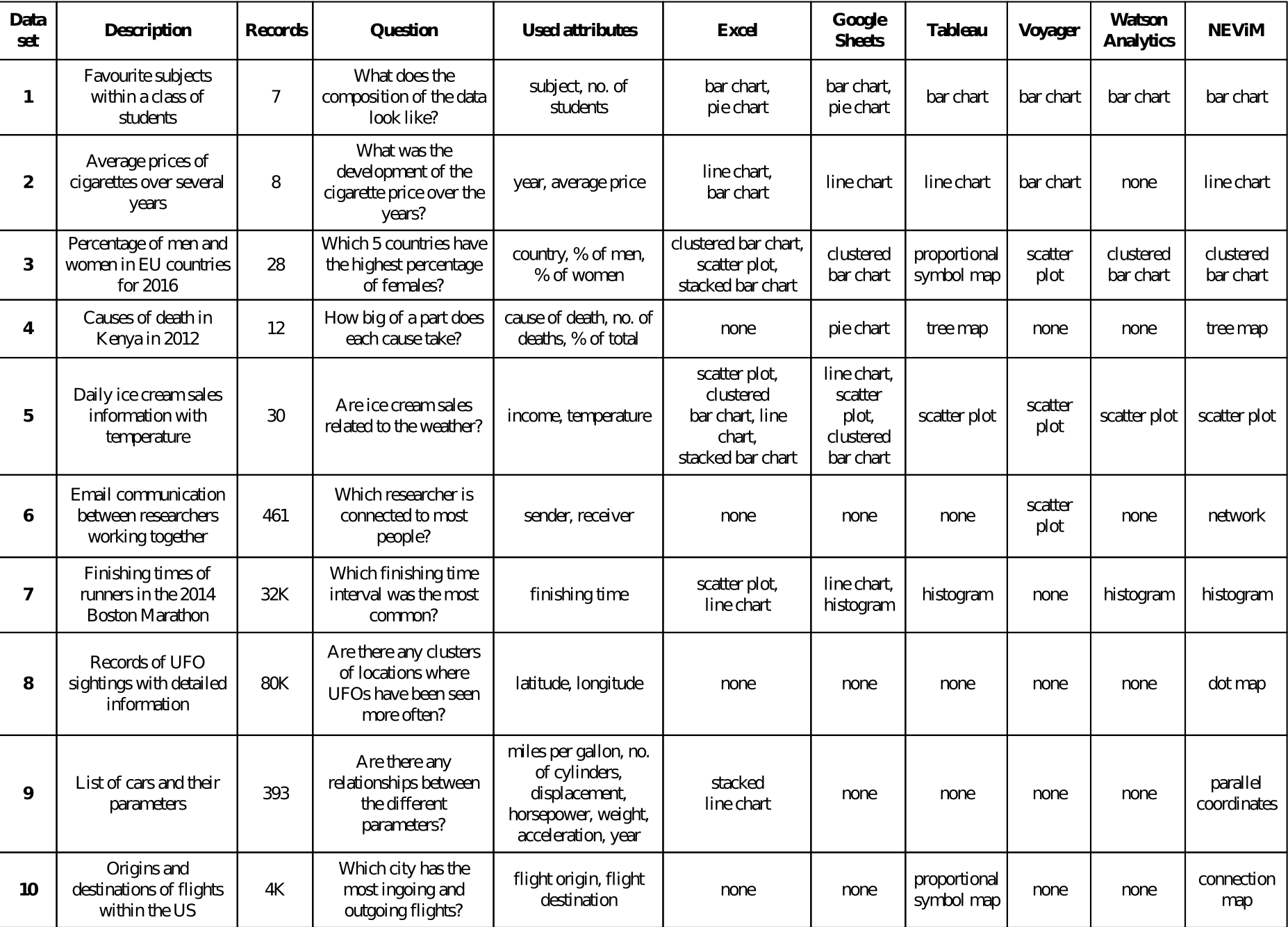}
\end{table*}

We named our model \textbf{NEViM}, which stands for Non-Expert Visualization Model. We tested our model against existing solutions which are freely available: Tableau (10.1.1), Watson Analytics (version available in July 2017), Microsoft Excel (15.28 Mac), Voyager (2) and Google Sheets (version available in July 2017). For each system and every data set, we aimed to achieve a recommendation for a data visualization that would answer the question and incorporate all the specified attributes in one graph as there is no possible way to answer the question without incorporating the specified attributes. Some systems solve more complex questions by creating a series of different data visualizations, with each visualization incorporating a different combination of attributes. We excluded such solutions from our test results because we feel that it is a workaround. For Microsoft Excel and Google Sheets, the recommendation process results in several recommendations and the systems do not rank them. For these cases we recorded all valid recommendations. 

\subsubsection{Results}

For data set 1, all systems recommended a bar chart. Excel and Google Sheets also recommended a pie chart. The recommendations for data set 2 were either line charts or bar charts. The specified question could be answered by either of these. Watson Analytics was not able to give a recommendation because it could not recognize that the average price attribute was a number. We have attempted resolving this issue but were not able to. For data set 3, the majority recommendation was a clustered bar chart, in line with the recommendation made by NEViM. Data set 4 proved to be challenging for Voyager and Watson Analytics. Since the data was hierarchical and the question was asking to see parts-of-whole, a suitable solution would be a tree map. A pie chart shows parts-of-whole, but does not indicate hierarchy. The question asked for data set 5 could be answered using different types of data visualizations. Since it is asking to analyze the correlation between 2 variables, a scatter plot is a suitable solution. All systems recommended it. Data set number 6 was an example of a social network, thus the most suitable visualization would be a network. However, the answer to the specified question could also be answered with a scatter plot as suggested by Voyager. This is because networks can also be represented as adjacency matrices and the scatter plot generated by Voyager is essentially an adjacency matrix. Data set 7 and its question were aimed at visualizing distributions. Distributions can be visualized, among others with histograms, scatter plots and line charts. Data set 8 was an example of spatial data. Spatial data is best visualized through maps. Tableau offers map visualizations but we suspect that it cannot plot on the map according to latitude and longitude coordinates. Watson Analytics and Google Sheets have the same issue. Microsoft Excel and Voyager do not support maps at all. In Data set 9 the answer to the question was revealed through comparing 7 attributes. This meant that the visualization has to support 7 different variables. Both stacked line chart and parallel coordinates are valid solutions. The final data set 10 was again spatial. This time it could be solved through plotting on a map but also by analyzing the distribution of the data set. Both proportional symbol map and connection map (as a flight implies a connection between two cities) are valid solutions.

\subsection{Extendibility test}

\begin{figure}
\centering\includegraphics[width=0.5\textwidth]{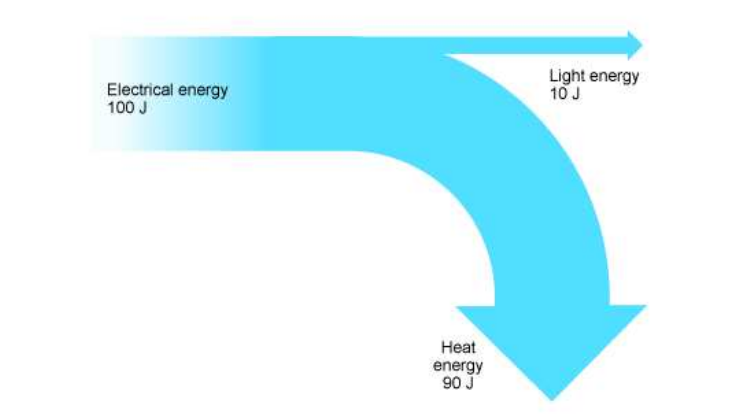}
\caption{Example of a Sankey diagram showing the distribution of energy in a filament lamp \cite{bbc}} \label{sankey}
\end{figure}

The aim of this test was to determine whether our model was easily extendible. We went through the process of adding a new data visualization type - a Sankey diagram. Sankey diagrams are flow diagrams that display quantities in proportion to one another. An example of a Sankey diagram can be seen in Figure \ref{sankey}. 

\noindent We look into the classifications that we already have and search for the most similar one. We find out that the Tree Map has the same classification, so we need to find a distinguishing feature between a Tree Map and a Sankey diagram. That feature is, that a Sankey diagram shows flow. We search through the model and find occurrences of a Tree Map. We then add a question asking "Do you want to show flow?". If the user answers "Yes", he gets a recommendation for a Sankey diagram. If he answers "No" he gets a Tree Map.

\begin{figure*}[!ht]
  \includegraphics[width=\textwidth]{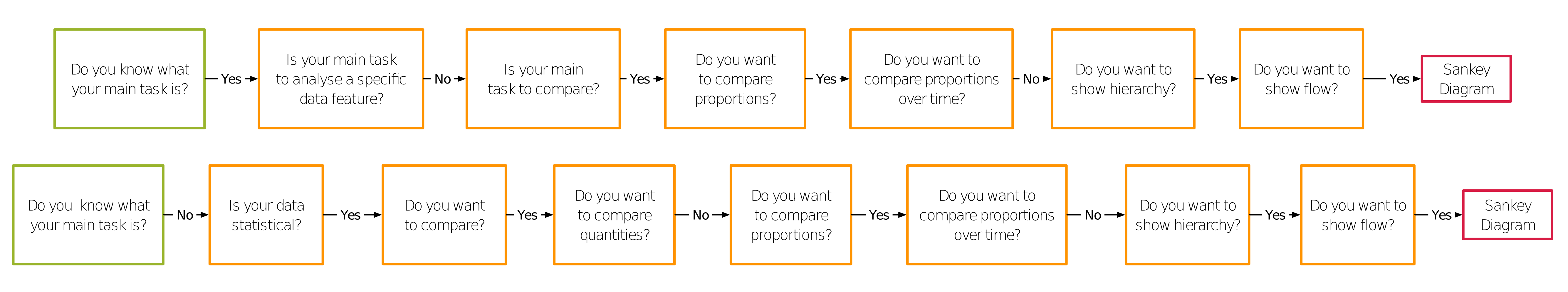}
  \caption{Two possible paths to reach a Sankey diagram (left: task-based, right: data-based) \cite{nevim}}
  \label{fig:paths}
\end{figure*}

\section{Evaluating the model}

\subsection{Test results} \label{advantages}

For the competitivness test, we can observe that NEViM provided usable solutions in all cases. The users have several paths that they can take through NEViM to get to a recommendation, depending on what information they know about their data or their task. NEViM has an advantage that it is not limited by implementation. Since two of our data sets were aimed at spatial data visualization and one at network data visualization, some systems were not able to make recommendations simply because they do not support such visualization types. Furthermore, NEViM includes more types of visualizations than any of the current systems, which results in recommendations for specialty visualizations that can be more suitable for a certain task. Another advantage is that it always results in only one recommendation, unlike Microsoft Excel or Google Sheets, where the user has to choose which one out of the set of recommendations to use. According to our survey, the most used visualization tool which incorporates a recommender system is Tableau (28\% of non-expert respondents). From the result table, we can see that in 5 out of 7 valid cases, NEViM made the same recommendation as Tableau. Furthermore, in data set 3 Tableau also made a recommendation for a Clustered Bar Chart, like NEViM did, but it was not the resulting recommendation. One of the attributes was the name of a country, so Tableau evaluated the data as spatial. We have noticed that whenever there is a geographical attribute, Tableau prefers to recommend maps, even though they might not be the most suitable solution. 

In the extendibility test we proved that it is indeed possible to add a new data visualization into the model. The process turned out to be easy, although we must acknowledge, that the finding of a distinguishing feature might be more complex in some cases. Before adding a visualization, one must be sure that they are adding a completely new type and not just a variation of an already existing visualization. This could potentially be very hard to determine, even for experts. We believe that the classification system we have adopted has potential to help with this. New distinguishing features could be added to the classification hierarchy, for example incorporating perceptual qualities of visualizations, their suitability for a specific field (such as finance) etc. You can read more about our ideas in Section \ref{future}.

\subsection{Fulfilling requirements}

\begin{enumerate}
\item \textbf{Simplicity} - Thanks to its question-based structure, using the model is simple. The user only has to answer yes or no questions. The basic structure is very straightforward. 
\item \textbf{Clarity} - The result of our recommendation system is a single data visualization, making it very clear. We believe that non-expert users need a clear answer to their visualization problem. If they are given a choice between two or more visualizations in the end, we believe that we have failed at the task of recommending them the most suitable one. We have narrowed their choices, but still have not provided a clear answer. However, this decision seems to be a controversial one, so it definitely needs to be validated through a user study. In the case that none of the data visualizations within the model are determined as suitable, the model still makes a recommendation to visualize using a table. 
\item \textbf{Versatility} - NEViM combines two different types of data visualization recommendation systems as defined in \cite{kaur}: task-oriented and data characteristics-oriented. These two types are distinguished by two different starting points within our model. Thanks to its base structure the model can be easily implemented in various different programming languages and environments.
\item \textbf{Extendibility} - To illustrate the extendibility of the model, we have added the Sankey diagram visualization. This proved to be a doable task. 
\item \textbf{Education} - This requirement has not been met yet. We feel that the fulfilment is more related to the implementation phase than the conceptual phase where we find ourselves now.  
\item \textbf{Transparency} - The traversal through our model is logical enough that it is clear why a certain type of data visualization was recommended. 
\item \textbf{Self-learning} - Our model is machine learning friendly and techniques can be applied for it to be able to self-learn.
\item \textbf{Competitiveness} - Through testing we have proved that our model produces recommendations similar or identical to existing solutions. It provided suitable solutions for all cases tested, unlike existing solutions.  
\end{enumerate}

\subsection{Advantages and disadvantages}

The advantages already been discussed in Section \ref{advantages} so we will not repeat them. A possible disadvantage of NEViM could be that the user has to either know what their main task is, or know what type of data they have. The question is, whether non-expert users will be able to determine this. We believe that this could be fixed through user testing to validate the overall structure of the model as well as the quality of the questions. The questions could be checked by a linguistics expert to see whether the wording is suitable and does not lead to possible ambiguous interpretations. 

Another disadvantage might lie in the fact that since we use data science terminology in our questions, we risk that non-experts might not be familiar with it and might not be able to answer the question. A solution could be to clarify the terms using a dictionary definition, which could pop up when the user hovers over the unfamiliar term. The solution is more part of the implementation phase, not the theoretical phase which we discuss here.

We have questioned whether the choice to recommend a table when no other suitable visualization is found is the correct one. There is an ongoing debate about when it is best to not visualize things, as discussed by Stephanie Evergreen \cite{evergreen}. Within the implementation phase, data could be collected to find out in how many cases the Table option is reached, to identify whether it is necessary to further address this issue.

\section{Implementing the model}

Earlier in this paper we have set a requirement for our model which stated that the model has to be easily implemented in different programming languages and environments. We will now implement the model we have constructed in a way that would allow it to be tested with users and adjusted. Since we are not very skilled in implementing a decision tree, we opted to build a prototype in a prototyping tool instead.

\begin{center}
\begin{figure*}[!ht]
  \centering\includegraphics[width=0.3\textwidth]{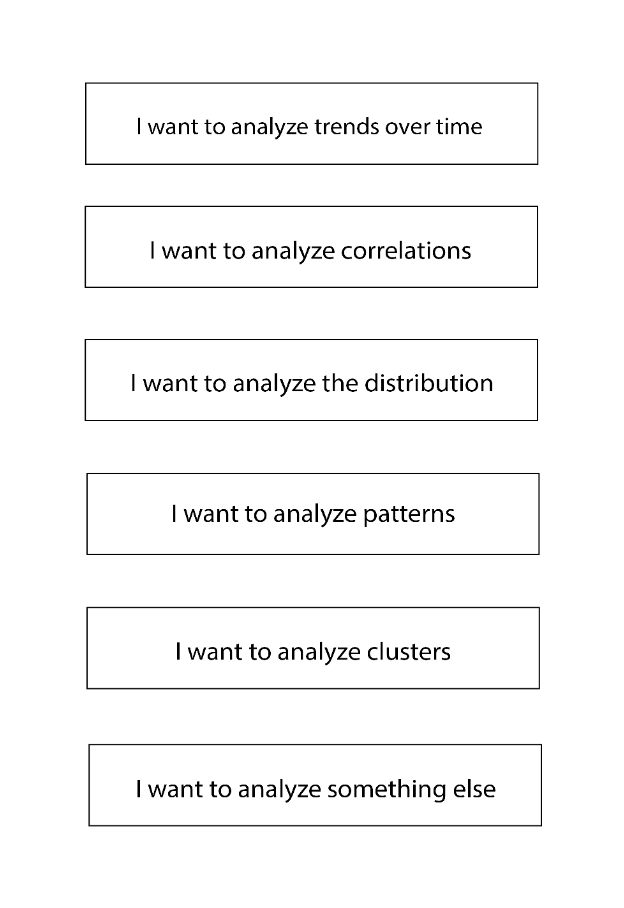}
  \caption{Example of a problematic screen from the prototype}
  \label{fig:prototype}
\end{figure*}
\end{center}

Once we found a suitable prototyping tool for our task. We set out to create an application, where each screen would be a question from our model. The user would then select either 'yes' or 'no' and be taken to the next question according to our model. This was an easy task, but very time consuming. Our final model had 107 internal nodes and we had to create a screen for each one. Luckily, we were able to reuse some of the screens as some of the questions occur in our model multiple times (for example once in the task-based branch and once in the data-based branch). When we finished this process, we realised another issue that our model has. The traversal through it could get very lengthy. The longest route from the root node to a leaf node was 12 questions. We came up with a way to solve this problem. We introduced the possibility of multiple choice answers rather than just 'yes' or 'no'. In Figure \ref{fig:prototype} you can see an example of a problematic screen from the prototype. In the first version of our prototype, we would have to go through the displayed options one by one, asking the user a question about each option. By introducing the solution displayed, we were able to shorten the lengthy routes and also remove possible bias of the user stemming from the order in which the questions were asked in the initial version. This solution also has another advantage in that the user can see a summary of the options in advance and it can help him in deciding what his main task is. Lastly, during this phase we also realised that we must provide the user with a fallback solution. What if the user simply does not know what their main task is? Which option would he select? We need to account for such an option in our model.

\section{Discussion \& Conclusions} \label{conclusion}

In this study we have managed to define, specify and go through the process of building a data visualization recommender system. We evaluated our approach by constructing our very own data visualization recommender system from scratch. We studied existing solutions thoroughly, we conducted a survey, we put together requirements for a model, we constructed a model, we tested the model, evaluated it and then made an example implementation. Initially we aimed for our model to be suited for non-expert users, but we were surprised to see that our survey showed that there is little to no difference in the attitude of non-experts and other users towards data visualization recommender systems. Throughout the paper we still comment on ways to make the model more suitable for non-experts, but because of the results of the survey, it became a secondary focus for us. 

We have proven that there is definitely a place for data visualization recommender systems in the data science world. We are very pleased with the model that we have constructed during the study and also with the fact that this paper can serve as a step-by-step guide for someone who wants to make a data visualization recommender system.

\section{Future Work} \label{future}

In the future, we would like to make use of the model that we have constructed as part of this study and further expand and improve it. We would like to perform more tests with more data sets and also carry out a usability test with different types of users. An idea could be to implement the model as a web application where users could rate the resulting recommendations, suggest new paths through the model or request new visualization types to be included. This would also validate the question paths that we have designed. The final recommendation could be enhanced with useful information about the data visualization type, tips on how to construct it, which tools to use and examples of already made instances. This would transform the model into a very useful educative tool. 

Another possible extension to the model could be to add another view which would incorporate information about the domain that the user's data comes from. There are data visualizations that are more suited for a specific data domain than others. For example, the area of economics has special types of data visualizations that are more suited to exposing different economic indicators. This would make the model part of the domain knowledge oriented data visualization systems recommender systems category according to \cite{kaur}.

We could introduce different features that could influence the visualization ranking - e.g. perceptual qualities of different data visualization types. Now that we have established a successful base, the possibilities for further development are endless.

\section*{Acknowledgements}

Research supported by SGS grant No. SGS17/210/OHK3/3T/18 and GACR grant No. GA18-18080S.

%
%
%

\begin{thebibliography}{8}

\bibitem{d3}
Data-Driven Documents (d3.js), \url{https://github.com/d3/d3/wiki/Gallery}. Last accessed 4 Aug 2017

\bibitem{nevim}
Kubern\'{a}tov\'{a}, P., Friedjungov\'{a}, M., van Duijn, M.: Knowledge at First Glance: A Model for a Data Visualization Recommender System Suited for Non-expert Users. In:  Proceedings of the 7th International Conference on Data Science, Technology and Applications - Volume 1: DATA, INSTICC, SciTePress (2018) 208-219

\bibitem{oneil}
O'Neil, C., Schutt, R.: Doing Data Science: Straight Talk From The Frontline. O’Reilly Media, Sebastopol,CA (2014)

\bibitem{tukey}
Tukey, J. W.: Exploratory Data Analysis. Addison-Wesley, Reading,MA (1970)

\bibitem{kirk}
Kirk, A.: Data visualization: A handbook for data driven design. SAGE, London,UK (2016)

\bibitem{illinsky}
Illinsky, N., Steele, J.: Designing data visualizations: representing informational relationships. O'Reilly Media, Sebastopol,CA (2011)

\bibitem{munzner}
Munzner, T., Maguire, E.: Visualization analysis and design. CRC Press, Boca Raton,FL (2015)

\bibitem{kaur}
Kaur, P., Owonibi, M.: A Review on Visualization Recommendation Strategies. In: Proceedings of the 12th International Joint Conference on Computer Vision, Imaging and Computer Graphics Theory and Applications (2017) 266-273

\bibitem{bharat}
Gnanamgari, S.: Information presentation through default displays. Ph.D. dissertation, Philadelphia,PA,USA (1981)

\bibitem{apt}
Mackinlay, J.: Automating the design of graphical presentations of relational information. In: Acm Transactions On Graphics (Tog) 5.2 (1986) 110-141

\bibitem{casner}
Casner, S., Larkin, J.H.: Cognitive efficiency considerations for good graphic design. Carnegie-Mellon University Artificial Intelligence and Psychology Project, Pittsburgh,PA (1989)

\bibitem{roth}
Roth, S.F., Mattis, J.: Data characterization for intelligent graphics presentation. In: SIGCHI Conference on Human Factors in Computing Systems (1990)

\bibitem{vizql}
Hanrahan, P.: Vizql: a language for query, analysis and visualization. In: Proceedings of the 2006 ACM SIGMOD international conference on Management of data. ACM (2006)

\bibitem{stolte}
Stolte, C.: Polaris: A system for query, analysis, and visualization of multidimensional relational databases. In: IEEE Transactions on Visualization and Computer Graphics 8.1 (2002) 52-65

\bibitem{mackinlay2}
Mackinlay, J., Hanrahan, P., Stolte, C.: Show me: Automatic presentation for visual analysis. In: IEEE Transactions on Visualization and Computer Graphics 13.6 (2007)

\bibitem{manyeyes}
Viegas, F., Wattenberg, M., van Ham, F., Kriss, J., McKeon, M.: ManyEyes: a site for visualization at internet scale. In: IEEE Transactions on Visualization and Computer Graphics 13.6 (2007)

\bibitem{watson}
Smart data analysis and visualization, \url{https://www.ibm.com/watson-analytics}. Last accessed 4 Aug 2017

\bibitem{vizdeck}
Key, A., Perry, D., Howe, B., Aragon, C.: Vizdeck: self-organizing dashboards for visual analytics. In: Proceedings of the 2012 ACM SIGMOD International Conference on Management of Data (2012)

\bibitem{excel}
Available chart types in Office, \url{https://support.office.com/}. Last accessed 4 Aug 2017

\bibitem{seedb}
Vartak, M., Madden, S., Parameswaran, A., Polyzotis, N.: SeeDB: supporting visual analytics with data-driven recommendations. VLDB (2015)

\bibitem{voyager}
Wongsuphasawat, K., Moritz, D., Mackinlay, J., Howe, B., Heer, J.: Voyager: Exploratory analysis via faceted browsing of visualization recommendations. In: IEEE Transactions on Visualization and Computer Graphics 22.1 (2016) 649-658

\bibitem{compass}
Vega Compass, \url{https://github.com/vega/compass}. Last accessed 4 Aug 2017

\bibitem{vegalite}
Satyanarayan, A., Moritz, D., Wongsuphasawat, K., Heer, J.: Vega-lite: A grammar of interactive graphics. In: IEEE Transactions on Visualization and Computer Graphics 23.1 (2017) 341-350

\bibitem{googlesheets}
Chart and Graph Types, \url{https://support.google.com/}. Last accessed 9 Aug 2017

\bibitem{wehrend}
Wehrend, S., Lewis, C.: A problem-oriented classification of visualization techniques. In: Proceedings of the 1st Conference on Visualization '90, IEEE Computer Society Press (1990)

\bibitem{improvise}
Zhou, M. X., Feiner, S. K.: Visual task characterization for automated visual discourse synthesis. In: Proceedings of the SIGCHI conference on Human factors in computing systems, ACM Press/Addison-Wesley Publishing Co. (1998)

\bibitem{gotz}
Gotz, D., Wen, Z.: Behavior-driven visualization recommendation. In: Proceedings of the 14th international conference on Intelligent user interfaces, ACM (2009)

\bibitem{zhou2}
Zhou, M. X., Chen, M., Feng, Y.: Building a visual database for example-based graphics generation. INFOVIS 2002 IEEE Symposium (2002)

\bibitem{dataslicer}
Alborzi, F., Reutter, J., Chaudhuri, S.: DataSlicer: Task-Based Data Selection for Visual Data Exploration. arXiv preprint (2017)

\bibitem{bbc}
Bbccouk, \url{http://www.bbc.co.uk/schools/gcsebitesize/science/aqa/energyefficiency}. Last accessed 17 Aug 2017

\bibitem{evergreen}
Evergreen, S. D.: Effective data visualization: The right chart for your data. SAGE Publications, Thousand Oaks,CA (2016)

\bibitem{yau1}
Yau, N.: Visualize This: The FlowingData Guide to Design, Visualization, and Statistics. John Wiley and Sons, Hoboken,NJ (2011)

\bibitem{yau2}
Yau, N.: Data points: Visualization that means something. John Wiley and Sons, Hoboken,NJ (2013)

\bibitem{heer}
Heer, J., Bostock, M., Ogievetsky, V.: A tour through the visualization ZOO. Queue 8.5 (2010)

\bibitem{hardin}
Hardin, M., Hom, D., Perez, R., Williams, L.: Which chart or graph is right for you?. Tell Impactful Stories with Data. Tableau Software (2012)

\bibitem{yuk}
Yuk, M., Diamond, S.: Data visualization for dummies. John Wiley and Sons, Hoboken,NJ (2014)

\bibitem{brath}
Brath, R., Jonker, D.: Graph analysis and visualization: discovering business opportunity in linked data. John Wiley and Sons, Hoboken,NJ (2015)

\bibitem{borner}
Borner, K., Polley, D. E.: Visual insights: A practical guide to making sense of dat. MIT Press, Cambridge,MA (2014)

\bibitem{telea}
Telea, A. C.: Data visualization: principles and practice. CRC Press, Boca Raton,FL (2007)

\bibitem{borner2}
Borner, K.: Atlas of knowledge: Anyone can map. MIT Press, Cambridge,MA (2015)

\bibitem{ware1}
Ware, C.: Visual thinking: For design. Morgan Kaufmann, Burlington,MA (2010)

\bibitem{ware2}
Ware, C.: Information visualization: perception for design. Elsevier, Amsterdam,NL (2012)

\bibitem{stacey}
Stacey, M., Salvatore, J., Jorgensen, A.: Visual intelligence: Microsoft tools and techniques for visualizing data. John Wiley \& Sons, Hoboken,NJ (2015)

\bibitem{hinderman}
Hinderman, B.: Building responsive data visualization for the web. John Wiley \& Sons, Hoboken,NJ (2015)

\bibitem{gemignani}
Gemignani, Z., Gemignani, C., Galentino, R., Schuermann, P.: Data fluency: Empowering your organization with effective data communication. John Wiley \& Sons, Hoboken,NJ (2014)

\end{thebibliography}
%

\end{document}